# Two Approaches to the Identity of Processes in BFO

Fumiaki Toyoshima[1] and Adrien Barton[2]

[1] IRIT, Université de Toulouse, 118 Route de Narbonne, F-31062 Toulouse Cedex 9, Toulouse, France
[2] IRIT, CNRS, Université de Toulouse, 118 Route de Narbonne, F-31062 Toulouse Cedex 9, Toulouse, France

**Abstract**

This paper aims to explore processes and their identity with a focus on the upper ontology Basic Formal Ontology (BFO). We begin with a classification based on two basic classes of changes of independent continuants: changes with respect to a single specifically dependent continuant thereof or with respect to the spatial region that its parts occupy. We accordingly distinguish two kinds of simple processes: specifically dependent continuant changes and spatial changes. Next, we investigate a compositional approach to the identity of processes: the identity of any process is determined by the identity of the simple processes that compose them. Then, we consider a causal approach to the identity of processes with recourse to a dispositional view of processes according to which any process is a realization of some disposition. We also examine assumptions on which these two approaches to the identity of processes are based.

## 1. Introduction

The distinction between continuants (aka endurants) and occurrents (aka perdurants) is widely accepted in many upper/foundational ontologies. The basic idea is that continuants are entities that persist in time and that can undergo changes, whereas occurrents are entities that unfold themselves in time and that can be changes of continuants. Examples of continuants include material objects (e.g. molecules, people, and planets) and properties in the broad sense of the term (e.g. the color of this apple and the fragility of glass). Paradigmatic examples of such occurrents are often grouped under the heading of "process" or "event": cell division, the life of this person, and the earth orbiting around the sun, for instance. There is a growing demand for a solid ontology of such occurrents, as is illustrated by the fact that, besides molecular function and cellular component, biological process is one of the three principal categories in the Gene Ontology (GO) [1].

In this paper we will explore an ontology of processes compatible with the foundational framework of Basic Formal Ontology (BFO) [2][3][4]. We use the term "process" in the BFO sense of the term throughout the paper (see Section 7 for a radically different view of processes and events from BFO's). One reason why we investigate the BFO ontology of processes is that it may remain relatively underspecified up to date. For example, process profiles [5] have been before proposed in the BFO community: two heart beating processes of the "same rate" can be analyzed as having as parts two instances of the same process profile universal such as *72bpm rate process profile*. But they have been eventually left out of the latest BFO version [3]. For a noteworthy recent work on processes in BFO, Jarrar & Ceusters [6] propose a classification of processes in BFO by focusing on how some well-known aspectual notions used to classify verbal phrases — viz. homeomericity, cumulativity, telicity, instantaneity, and atomicity — can be ontologically reinterpreted to build BFO-based process ontologies. It will be a valuable complementary study for us to consider carefully what kinds of changes processes are and on which conditions one process is the same as another.

The paper centers around the fundamental question of what is the identity of processes, or what is a set of necessary and jointly sufficient conditions (represented by a "if and only if" or "iff" clause) for two processes being identical.[2] It is organized as follows. Section 2 is devoted to preliminaries. On the basis of the recent work by Guarino, Baratella and Guizzardi [8] (henceforth "GBG"), Section 3 introduces two kinds of processes: specifically dependent continuant changes and spatial changes. Section 4 investigates a compositional approach to the identity of processes: it may be characterized with identity criteria for specifically dependent continuant changes and spatial changes, on the assumption that any process is a mereological sum of these two kinds of processes. Section 5 develops a causal approach to the identity of processes based on a dispositional view of processes according to which any process is a realization of some disposition. Section 6 offers discussion. Section 7 discusses related work. Section 8 concludes the paper.

## 2. Preliminaries
## 2.1. The basic structure of BFO

BFO is an upper ontology that is theoretically underpinned by the realist methodology for ontology development [9], according to which ontologies should represent actual entities as described by science, as well as by perspectivalism: BFO is perspectival along two major dimensions, of continuants and occurrents and these dimensions may provide equally accurate descriptions of the same reality. Continuants persist in time: they maintain their identity and may gain or lose parts over the course of time. Occurrents unfold themselves through time. (Note that we will assume the framework of classical physics in this article as BFO often does, even if there are plans to extend BFO beyond the classical realm.)

Continuants are further divided into independent continuants (such as material objects and spatial regions) and dependent continuants. Among dependent continuants are specifically dependent continuants, which depend (existentially) on at least one independent continuant. Two major subtypes of specifically dependent continuants are realizable entities and qualities. The former can be realized in processes of specific correlated types in which the bearer participates: e.g. the disposition of this glass to be broken, the function of this heart to pump blood, and the role of being a doctor (for more thoughts, see Toyoshima et al.'s [10] systematic study of realizable entities in BFO). They can be present even when not realized: this glass is fragile even if it is not broken, for instance. The latter are fully exhibited or manifested or realized if they are borne at all: e.g. color, shape, and mass.

Among realizable entities, a disposition in BFO is defined as: "A realizable entity (…) that exists because of certain features of the physical makeup of the independent continuant that is its bearer. One can think of the latter as the material basis of the disposition in question" ([2], p. 178). Typical examples of dispositions include fragility (the disposition to break when pressed with sufficient force) and solubility (the disposition to dissolve when put in a solvent). The material basis of a disposition is some material part(s) of the disposition bearer in virtue of which the disposition exists. BFO also describes a disposition as an "internally grounded realizable entity": if a disposition ceases to exist, then the physical makeup of the bearer is thereby changed. The fragility of this glass has as material basis some molecules of the glass and the glass is physically changed when it is no longer fragile, for instance.

As for occurrents, a process is an occurrent that exists in time by occurring, has temporal parts, and depends on at least one independent continuant as participant. A spatiotemporal region is an occurrent at or in which occurrent entities (notably processes) can be located. A temporal region is an occurrent that results from the projection of a spatiotemporal region onto the temporal dimension (for more thoughts, see Galton's [11] discussion on temporal and spatiotemporal regions in BFO).

---

[2] We take it for granted that identity criteria play an important role in foundational ontology research. For deeper thoughts, see Garbacz's [7] philosophical argument that identity criteria can help to specify "ways" or "modes" in which identity facts hold.

## 2.2. Ontology of dispositions

We will use a BFO-compliant extended theory of dispositions along with previous works [10][12][13][14]. To be realized in a process, a disposition needs to be triggered by some other process: a process of pressing this glass with sufficient force triggers the fragility of the glass, which is realized in a process of glass-breaking. Note that dispositions may exist even if they are not realized or even triggered: a glass is fragile even if it never breaks or even if it never undergoes any shock. We will also utilize what Barton et al. [14] call the "PARTHOOD" model of dispositions, according to which a part of a realization of a disposition is also a realization of this disposition: for instance, the short cracking process of this glass that immediately precedes its splitting into many pieces is a realization of the fragility of the glass because it is part of the glass-breaking process (in which the fragility is realized).

## 2.3. Categories and relations

We will introduce the terms for BFO categories and their associated unary predicates — see the taxonomy depicted in Figure 1 (where a type *A* being a subtype of a type *B* implies all instances of *A* being instances of *B*). We will also introduce the terms for relations and their associated relational predicates — see Table 1 for a list of relational predicates and their explanation. As for parthood, we will assume so-called classical (extensional) mereology (e.g. [15], Section 2).

In formalization, variables and individual constants stand for particulars, predicates stand for universals and defined classes (unary predicates) and relations, and free variables are universally quantified. We will employ conventional logical symbols of first-order logic with identity. In the text, terms for instances and classes will be boldified and italicized, respectively: for example, this particular person **John** and the human type *Human*.

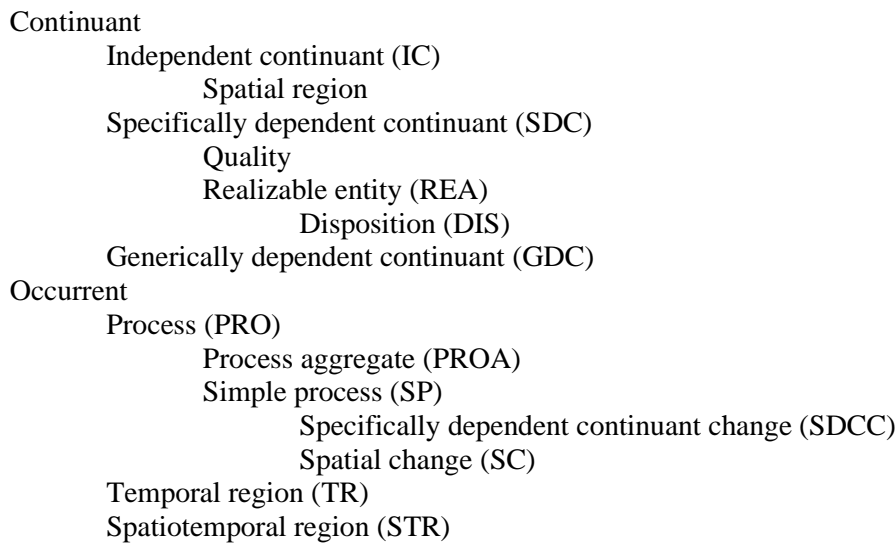

Continuant
    Independent continuant (IC)
        Spatial region
    Specifically dependent continuant (SDC)
        Quality
        Realizable entity (REA)
            Disposition (DIS)
    Generically dependent continuant (GDC)
Occurrent
    Process (PRO)
        Process aggregate (PROA)
        Simple process (SP)
            Specifically dependent continuant change (SDCC)
            Spatial change (SC)
    Temporal region (TR)
    Spatiotemporal region (STR)

**Figure 1**: Taxonomy of BFO categories and their associated unary predicates (categories added by us are underlined)

**Table 1**
List of relational predicates, their domains, ranges and semantic reading, and their functionality.

| Relational predicate | Domain, range, and semantic reading | Functionality |
|---|---|---|
| INH(x,y) | x (SDC) inheres in y (IC) | Functional |
| OSTR(x,y) | x (PRO) spatiotemporally occupies y (STR) | Functional |
| OTR(x,y) | x (PRO) temporally occupies y (TR) | Functional |
| P(x,y) | x is part of y | / |
| PC(x,y,t) | x (IC)[3] participates in y (PRO) at t (TR) | / |
| PCSP(x,y) | x (IC) participates in simple process y (SP) | /[4] |
| PSDC(x,y) | x (SDCC) is a change of y (SDC) | Functional |
| REAL(x,y,t) | x (REA) is realized in y (PRO) at t (TR) | / |
| SUM(y,$x_1$, …, $x_n$) | y is a mereological sum of $x_1$, …, $x_n$ | Functional on y |

## 3. Simple processes

A process is, in nature, a change of some participant(s) of this process. GBG define *simple events* as qualitative changes by triples <*o,q,t*> where *o* is the object of change, *q* is the subject of change — which is a quality, as found in the Descriptive Ontology for Linguistic and Cognitive Engineering (DOLCE) [16][17] — and *t* is the time during which the change happens. Either *q* inheres in *o*, in which case this is a *direct qualitative change*; or *q* inheres in a part of *o*, in which case this is an *indirect qualitative change*. For example, when a person gesticulates by moving his hand: the gesticulation is an indirect change of the person, whereas the hand moving is a direct change of his hand.

We will here endorse a similar view on the metaphysics of processes[5], applied to BFO:independent continuants: when an independent continuant changes, it always changes *with respect to some aspect* of that independent continuant. Note however that position is a quality in GBG's sense but not in BFO. Therefore, we need to introduce two kinds of simple processes involving what GBG call a direct qualitative change of an independent continuant: specifically dependent continuant (SDC) changes, namely change with respect to a single specifically dependent continuant of the independent continuant (presented in Section 3.1); and spatial changes, namely change with respect to the spatial regions that its parts occupy (presented in Section 3.2). We will deal in this section with direct qualitative change in the sense of GBG; and come back in Section 6.3 to a possible reduction of indirect qualitative change to direct qualitative change. For an illustrative purpose, we will employ Davidson's [18] famous example in which a sphere $s_1$ is rotating and heating at the same time.

### 3.1. Specifically dependent continuant change

In the driving example, we can identify this process **$p_{heat}$** of $s_1$ heating up such that **$p_{heat}$** has $s_1$ as participant. Then **$p_{heat}$** is a change of $s_1$ with respect to the temperature, say **$temperature_1$** — which is a quality — of $s_1$. Suppose that $s_1$ is at 60 degrees Celsius at time $t_1$ and at 70 degrees Celsius at time $t_2$ such that **$p_{heat}$** temporally occupies a connex temporal region encompassing $t_1$ and $t_2$.[6] Following BFO's standard view ([2], p. 97), we introduce the term "determinate" for universals: roughly, a universal *X* is determinate of a universal *Y* if and only if being *X* is a specific way of being *Y* [19]. The received analysis of this case of temperature change appeals to two determinates of the determinable *Temperature*, namely *60°C Temperature* and *70°C Temperature*: **$temperature_1$** (which is an instance of *Temperature*) is an instance of *60 °C Temperature* at time $t_1$ and an instance of *70 °C Temperature* at time $t_2$.

---

[3] Note that for BFO [4], SDCs and GDCs can also participate in processes, but we do not investigate here the details of such participation relations.
[4] See Footnote 7 for some remark about the possible inverse functionality of PCSP.
[5] Our focus will be here purely ontological and not semantic: we do not enter into the question of how events can be described. Therefore, we will not discuss GBG's notions of focus, core context or internal context, that arguably belongs to the semantics of event descriptions.
[6] We are assuming that **$p_{heat}$** (or any other process instance appearing in this paper, unless otherwise specified) is a temporally continuous process, although there are also temporally discontinuous processes such as a football match interrupted by a blackout. See also Section 5.4 for the possible implications of temporally discontinuous processes for a dispositional view of processes (to appear in Section 5.1).

We can say that **p_heat** is a process in which **s_1** changes with respect to **temperature_1**. The analysis of this example can lead to the following definition of the term "specifically dependent continuant change" such that **p_heat** is a specifically dependent continuant change:

> specifically dependent continuant change
> =_def. A process that is a change of an independent continuant with respect to a single specifically dependent continuant thereof.

SDC changes can concern qualities such as temperature, but also realizable entities — for example, a metal sphere becoming more or less ductile (where ductility is a disposition to change shape). They can also concern the coming into existence of an SDC (e.g. the appearance of transparency of a portion of sand as it is transformed into glass) or the ceasing to exist of an SDC (e.g. the disappearance of structural integrity of a window as it is broken). We will henceforth employ the expressions, such as "**p_heat** is a change of **temperature_1**" (formally: PSDC(**p_heat**, **temperature_1**)), to characterize specifically dependent continuant changes.

We can then formalize the definition of a specifically dependent continuant change in terms of PSDC, which is taken to be functional:

> **D1** SDCC(p) =_def. PRO(p) ∧ ∃sdc PSDC(p,sdc)
> "*p* is a specifically dependent change" means: *p* is a process and there exist *sdc* such that *p* is a change of *sdc*.

## 3.2. Spatial change

In our driving scenario, we can identify this process **p_rot** of **s_1** rotating such that **p_rot** has **s_1** as participant. Then **p_rot** is a change of **s_1** with respect to the spatial region that its parts occupy: parts of **s_1** occupy different spatial regions at different times over the course of **p_rot**. We propose the following definition of the term "spatial change" such that **p_rot** is a spatial change:

> spatial change =_def. A process that is a change of an independent continuant with respect to the spatial region that some part thereof occupies.

What is commonly called "motion process" is a spatial change, although we might imagine more exotic forms of spatial changes (e.g. teleportation).

## 3.3. Discussion of simple processes

Note that the categories of Spatial change and SDC change are not disjoint. For example, a change of shape of a sponge as I press it is a spatial change (as its parts are moving through space), but also arguably an SDC change, since shape is commonly considered as a quality (and thus an SDC); indeed, examples of qualities in BFO include "the shape of this hand" ([2], p. 96).

Note also that simple processes are not all atomic: some simple processes can have as proper part other simple processes. For example, the spatial change of rotation of the sphere has as part the spatial changes of rotation of its upper hemisphere and of rotation of its lower hemisphere. Similarly, the color change of an apple from green to red has as part the color changes of its left half and of its right half.

## 4. Compositional approach to the identity of processes

We first investigate a compositional approach to the identity of processes. The central idea is that a general criterion of the identity of process aggregates (that are sums of specifically dependent continuant changes and spatial changes) can be provided in terms of the identity criteria for specifically dependent continuant changes and spatial changes. This would provide a criterion of identity of processes if one hypothesizes that any process is a mereological sum of instances of those two kinds of changes.

## 4.1. Identity criterion for specifically dependent continuant changes

The identity condition of a specifically dependent continuant change can be provided in terms of the relevant specifically dependent continuant(s) therein and the time at which it occurs:

**A1** $PSDC(p_1,sdc_1) \land PSDC(p_2,sdc_2) \land OTR(p_1,t_1) \land OTR(p_2,t_2) \rightarrow$
$[p_1=p_2 \leftrightarrow (sdc_1=sdc_2 \land t_1= t_2)]$
If $p_1$ is a change of *sdc₁*, $p_2$ is a change of *sdc₂*, $p_1$ temporally occupies *t₁*, and $p_2$ temporally occupies *t₂*, then: *p₁* is identical with *p₂* iff *sdc₁* is identical with *sdc₂* and *t₁* is identical with *t₂*.

For instance, the identity of **p_heat** is determined by the quality **temperature₁** and the temporal interval which **p_heat** occupies.

## 4.2. Identity criterion for spatial changes

The identity condition of a spatial change can be provided in terms of its participants[7] and the time at which it occurs:

**A2** $SC(p_1) \land SC(p_2) \land OTR(p_1,t_1) \land OTR(p_2,t_2) \rightarrow$
$[p_1=p_2 \leftrightarrow [\forall x(PCSP(x,p_1) \leftrightarrow PCSP(x, p_2)) \land t_1= t_2)]]$
If $p_1$ is a spatial change, $p_2$ is a spatial change, $p_1$ temporally occupies *t₁*, and $p_2$ temporally occupies *t₂*, then: *p₁* is identical with *p₂* iff 1) for any *x*, *x* participates in the simple process *p₁* iff *x* participates in the simple process *p₂* and 2) *t₁* is identical with *t₂*.

For example, the identity of **p_rot** is determined by its participants (in particular **s₁**) and the temporal interval during which **p_rot** occurs.

## 4.3. Identity criterion for process aggregates

Then, we introduce the term "process aggregate" that can be defined in natural and formal languages as follows, although it is formally inexpressible in first-order logic owing to the use of the natural number *n* (which will also apply to A3 below):

process aggregate =_def. A process that is a sum of multiple different simple processes.

**D2** $PROA(p) =_{def.} PRO(p) \land \exists n, sp_1,...,sp_n (n \geq 2 \land \bigwedge_{1 \leq i \leq n} SP(sp_i) \land SUM(p, sp_1,...,sp_n) \land sp_1 \neq sp_2)$
"*p* is a process aggregate" means: *p* is a process, and there are at least two different simple processes *sp₁,…,spₙ* such that *p'* is a sum of *sp₁,…,spₙ*.

Note that a process aggregate can be just a sum of specifically dependent continuant changes, just a sum of spatial changes or a sum of some SDC change(s) and some spatial change(s). To illustrate process aggregates with our driving example, we can think of a process aggregate that is the sum of the spatial change **p_rot** and many specifically dependent continuant changes (such as **p_heat**).

---

[7] Adapting GBG's conception of participation, we might want to formulate the relation of independent continuant participation generally as follows. An independent continuant *x* participates in a process *p* just in case: 1) If *p* is a simple process, then *x* is the mereologically maximal independent continuant changing in *p*; or 2) If *p* is a process aggregate, then *x* is an independent continuant changing in one of simple processes that are parts of *p*. In particular, a simple process would have only one participant. However, more investigation is required to check whether this would fit with BFO's pre-formal characterization of participation: we might want to say that not only x, but also parts of x participate in a process of motion of x.

We can also provide the identity condition of process aggregates. Informally speaking, two process aggregates are identical iff they have as parts the same simple processes. To put it formally:

> **A3** $[\text{PROA}(p_1) \wedge \text{PROA}(p_2) \wedge p_1 = p_2] \leftrightarrow \exists n, sp_1, ..., sp_n,$
> $(n \geq 2 \wedge \bigwedge_{1 \leq i \leq n} \text{SP}(sp_i) \wedge \text{SUM}(p_1, sp_1, ..., sp_n) \wedge \text{SUM}(p_2, sp_1, ..., sp_n))$
>> Two process aggregates $p_1$ and $p_2$ are identical iff there exist simple processes $sp_1, ..., sp_n$, such that both $p_1$ and $p_2$ are the sum of $sp_1, ..., sp_n$.

## 4.4. Hypothesizing a general criterion for the identity of processes

Let us formulate the following hypothesis:

> **Process Decomposition Hypothesis (PDH)**
> Any process is a simple process or process aggregate, i.e. a mereological sum of simple processes (namely SDC changes and spatial changes). (Formally: $\text{PRO}(p) \leftrightarrow (\text{PROA}(p) \vee \text{SP}(p))$.)
> In particular, the PDH implies that the identity criteria for simple processes and for process aggregates provides an identity criterion for processes in general.

If the PDH is valid, then we have provided a general criterion for the identity of processes through A1, A2, and A3. This hypothesis seems to make sense on at least a variety of examples. For example, a dinner might be analyzed as a process aggregate composed by the motion of various utensils, food and body parts; some quality changes of the food; etc. To take another example, an apple rotting is arguably a process aggregate composed by processes of the color of the apple changing, its chemical composition changing, etc.

## 4.5. Discussion of the PDH

There are at least three kinds of processes (the two formers being identified by GBG [8]) that need to be consider to evaluate the cogency of the PDH: substantial change, e.g. a statue being created or destroyed; mereological change, e.g. a human gaining a tumor or losing a finger; and generically dependent change, e.g. a change of a document.

For each of those apparent changes, there is a variety of possible positions. One could endorse an eliminativist position, claiming that e.g. substantial change does not in fact exist. Alternatively, one could endorse a strong reductionist position, claiming that e.g. substantial change exists, but is in fact identical to some simple process or process aggregate. One could also endorse a weak reductionist position, claiming that e.g. substantial changes exist, and are based on simple processes that are "prior" or "more basic" [20], without being identical to such simple processes or their aggregates. Finally, one could endorse a non-reductionist position, claiming that e.g. substantial changes exist and are not based on more basic simple processes: they are all on equal ontological footing.

An eliminativist or strong reductionist position about all those three kinds of processes would not falsify the PDH; however, a weak reductionist or non-reductionist position of any of them would falsify it. Let us thus examine in turn each of those three changes.

### 4.5.1. Substantial change

Substantial change is classically analyzed as the coming into existence or the ceasing to exist of a substance — that is, in BFO, of a material entity. However, it is not clear that BFO should accept such changes. Indeed, BFO aims at being non-multiplicativist, as it states that no two material entities can occupy the same spatial region [3]. To use a canonical example, if an amount of clay has the shape of a statue, BFO does not distinguish two entities, the amount of clay and the statue, but rather considers that a unique substance, the amount of clay, plays the role of being a statue [21]. To take another example, a sand castle should not be distinguished from the collection (or mereological sum) of sand grains that constitute it. But then, if this sand castle is washed away by the sea, it might not imply in BFO that a substance (the sand castle) disappears. Rather, one might consider that the collection (or sum) of sand grains that was shaped in a castle-shape (and that instantiated the *Sand castle* class) is now scattered (and thus does not instantiate *Sand castle* anymore). Many other examples (a body rotting, a house being destroyed, etc.) might be similarly analyzed. Such a framework might imply an eliminativist view towards substantial change at the macroscopical scale: what exists is not a creation or destruction of substance[8], but rather a change in instantiating various classes by a material entity. This is not the only possible view though: BFO might also accept that an entity appears and disappears when the sand castle is created or destructed, although this may lead to some form of non-multiplicativism (since the sand castle and the aggregate of sand grains arguably do not have the same identity conditions and are thus distinct entities).

If the latter analysis is correct, then there are indeed processes of creation or destruction of material entities, which are arguably not reducible to simple processes, and thus the PDH is false. If the former analysis is correct, then the question for the validity of the PDH becomes whether a change in material entity instantiation can be reducible to a simple process or to a sum of simple processes. One might have an abundant view of SDCs that accepts SDCs such as "being a Sand Castle" (maybe as a sum of several more basic SDCs such as "being made of sand" and "having a castle-shape"). In this case, ceasing to instantiate *Sand Castle* would amount to the disappearance of this SDC, and such cases of substantial changes would be reducible to SDC change. In case BFO would reject such SDCs, though, we would need to add to the list of simple processes the change of instantiation by a material entity for the PDH to remain valid.

### 4.5.2. Mereological change

Let us now turn to mereological change: A process in which an independent continuant gains a part of loses a part. Consider e.g. the following processes:
- $p_{tpg}$: John gains a tumor in the pineal gland
- $p_{li}$: John loses his left index finger

With a sufficiently general conception of quality, we can account for such changes as simple changes. Suppose indeed that we accept the existence of the following qualities:
- $q_{tpg}$: John's quality of having a tumor in the pineal gland
- $q_{li}$: John's quality of having a full index finger

Then:
- $p_{tpg}$ is a simple change of $q_{tpg}$ (namely, its coming to existence)
- $p_{li}$ is a simple change of $q_{li}$ (namely, its ceasing to exist)

Therefore, the (strong or weak) reduction of mereological change to SDC change depends on whether BFO's understanding of SDCs is broad enough to accommodate SDCs such as $q_{tpg}$ and $q_{li}$ (consider also "being one-legged" or "having a mole on one's cheek").

### 4.5.3. GDC change

Finally, a generically dependent continuant (GDC) can arguably change. A GDC that is "dependent on one or other independent continuants and can migrate from one bearer to another" ([2], p. 179). An important example of GDCs are Information Content Entities (ICEs) [22], such as documents. It is an open question whether ICEs can change, but that seems possible: a document can, indeed, be filled or evolve — consider e.g. this article that evolved through time until its final state (see Barton et al.'s [23] discussion on some difficulties related to the diachronic identity of ICEs). Another kind of GDC might be social GDCs. Although those are not fully conceptualized in BFO (see Brochhausen et al.'s [24] work though), those might be an important kind of GDCs. For example, Arp et al. [2] consider that there is a social GDC that we might describe as corresponding to the role of President of the USA; Donald's Trump role of president of the USA (that existed from January 2017 to January 2021) and Joe

---
[8] In such a case, BFO might still want to accept the possibility of matter creation or annihilation at the microscopical scale, depending on what contemporary physics would have to say on this topic. We leave such considerations outside the scope of the present paper and stay in the realm of classical physics, as explained above.

Biden's role of president of the USA (that exists since January 2021) are two SDCs that might be concretizations of such a social GDC. In case a new law would change the power or responsibilities of the president of the USA, then this social GDC would arguably change.

All GDCs need to be concretized, often in SDCs (although some informational entities might be concretized in processes since BFO-ISO [3]). Thus, a GDC change might be seen as a parasitic entity over the change of the SDCs that concretize it, or over the processes that concretize it. However, BFO does not endorse an eliminativist approach of GDC; thus, it seems natural to consider that GDC changes should also not be eliminated. BFO also does not endorse a strong reductionist approach of GDC on their concretization: it does not identify a GDC with its concretization (or the sum of its concretizations). Therefore, it also seems natural to refrain from identifying a GDC change with, e.g. the change of the SDCs that concretize it. On the other hand, a weak reductionist (or even maybe non-reductionist) approach of GDC change would seem natural in BFO.

### 4.5.4. Conclusion for the PDH

Let us wrap up. Substantial change might be eliminated (but see Footnote 8) in favor of change of instantiation of a material entity, but it is an open question whether BFO would encompass a strong reductionist view of such latter changes. Mereological changes might be (strongly or weakly) reduced to quality changes. Finally, it does not seem that GDC changes can be eliminated or strongly reduced.

Thus, the PDH as formulated so far would be false. However, it might be saved by the combination of two moves: 1) endorsing a general enough view of SDC according to which change of material entity instantiation and mereological changes would be strongly reduced to SDC changes; and 2) widening the definition of simple processes in order to encompass not only SDC changes and spatial changes, but also GDC changes (and possibly material entity creations and destructions, in case BFO would accept such processes).

The second point implies in particular that we should spell out a criterion for the identity of GDC changes. A very straightforward criterion would then be a direct adaptation of the criterion (A1) proposed for SDCs above:

> *Axiom for GDC changes*
> $PGDC(p_1,gdc_1) \land PGDC(p_2,gdc_2) \land OTR(p_1,t_1) \land OTR(p_2,t_2) \rightarrow$
> $[p_1=p_2 \leftrightarrow (gdc_1=gdc_2 \land t_1= t_2)]$

## 5. Causal approach to the identity of processes
### 5.1. A dispositional view of processes

Since the compositional criterion of identity of processes proposed above crucially depends on the PDH, it would be nice to have another criterion of identity that would not rely on it. Thus, we next investigate a causal approach to the identity of processes. For this purpose, we will utilize a dispositional view of processes. The basic idea is that processes are entities that are causally brought about, and causation can be analyzed in terms of dispositions. For instance, Röhl & Jansen [12] maintain that: "dispositions connect the static structure of the world, i.e. the natural kinds of continuants, with the dynamical structure, i.e. the types of possible and actual causal processes" (ibid., p. 3). For that matter, the dispositional theory of causality has been actively developed in philosophical ontology [25][26].

One way to formalize such a dispositional view of processes is to hypothesize that any process is a realization of some disposition of an independent continuant that participates in that process:

> **A4** $PRO(p) \rightarrow \exists x,t,d(PC(x,p,t) \land REAL(d,p,t) \land INH(d,x))$
> For any process $p$, there exist $x$, $t$, and $d$ such that $x$ participates in $p$ at $t$, $d$ is realized in $p$ at $t$, and $d$ inheres in $x$.

To illustrate A4 with our driving example in the case of specifically dependent continuant changes, $p_{heat}$ is a realization of the disposition of $s_1$ to get heated. In the case of spatial changes, we could consider $p_{rot}$ as a realization of the disposition of $s_1$ to be realized in a process of rotational movement (cf. the view of Newtonian force as a disposition to be realized in a process of accelerated motion of the force bearer [27]).

## 5.2. A dispositional criterion for the identity of processes

Let us begin by considering two criteria that do not involve dispositions. One of the simplest criteria for the identity of processes is the identity of their participant(s) because a process depends on some independent continuant as a participant:

> **C1** Processes are identical iff they have the same participant(s) at any time.
> Formally: $PRO(p_1) \wedge PRO(p_2) \rightarrow [p_1 = p_2 \leftrightarrow \forall x,t\ (PC(x,p_1,t) \leftrightarrow PC(x,p_2,t))]$

Another criterion is the identity of the spatiotemporal regions of processes and it is traditionally popular in the philosophy of processes and events (as championed by Quine [28] and the late Davidson [29]):

> **C2** Two processes are identical iff they occupy the same spatiotemporal region.
> Formally: $PRO(p_1) \wedge PRO(p_2) \rightarrow [p_1 = p_2 \leftrightarrow \forall str\ (OSTR(p_1,str) \leftrightarrow OSTR(p_2,str))]$

But neither C1 nor C2 succeeds in identifying processes that we can intuitively differentiate. Using the driving example, we would otherwise have the consequence that $p_{heat}$ and $p_{rot}$ are both the same process by C1 (because they have the same participant, namely $s_1$) and by C2 (because they occur in the same place at the same time). This consequence may be undesirable in formal ontology as we may need to distinguish these processes when representing them in information systems.

Let us now turn to the criteria for the identity of processes that involve dispositions and their realizations. A straightforward dispositional criterion would be that two processes are identical iff they realize the same disposition(s) at the same time. We can formalize this statement as follows:

> **A5** $PRO(p_1) \wedge PRO(p_2) \rightarrow [p_1 = p_2 \leftrightarrow \forall d,t\ (REAL(d,p_1,t) \leftrightarrow REAL(d,p_2,t))]$
> Two processes $p_1$ and $p_2$ are identical iff: for any disposition $d$ and any temporal region $t$, $d$ is realized in $p_1$ at $t_1$ iff $d$ is realized in $p_2$ at $t$.

According to A5, for instance, $p_{heat}$ and $p_{rot}$ are both different processes because, as we have seen above, they are (albeit simultaneous) realizations of different dispositions of $s_1$: the disposition to get heated and the disposition to rotate, respectively.

## 5.3. Illustration: Clarifying BFO:History from a dispositional perspective

To illustrate the dispositional view of processes, we will analyze the subtype of *Process* called "*History*" in BFO, which is especially important, as this category enables us to define an injection from material entities (and sites) to processes. A BFO:*History* is: "A BFO: *process* that is the sum of the totality of processes taking place in the spatiotemporal region occupied by a material entity or site" ([2], p. 179). To be concrete, let us consider John's history from a dispositional viewpoint. "For example, the history of John is the sum of all processes that have occurred within John throughout the course of his entire life, at all granularities" (ibid., p. 123).

A naïve attempt to analyze John's history dispositionally would be to claim that it is the sum of *all* realizations of dispositions that inhere in any (proper or improper) part of John during his whole life. But this attempt fails because there exist some dispositions of John that are realized in processes that are not part of his history. Indeed, suppose that John is moving a pen at time $t_{move}$. This process $p_{move}$ is a realization of John's disposition $d_{John}$ to move something. The spatiotemporal region $str_{move}$ occupied by $p_{move}$ spatially projects onto the mereological sum of the spatial region occupied by John *and* the spatial region occupied by John's pen. Then $p_{move}$ is not part of John's history because John's history occupies only the spatiotemporal region occupied by John. Therefore, $d_{John}$ is a disposition of John that is realized in a process (namely $p_{move}$) that is not part of his history.

On closer examination, however, $d_{John}$ is also presumably realized in a pen moving-related process that *is* part of John's history. To see this, we will introduce Loebe's [30] notion of *processual role* in his theory of roles. A processual role is part of a process such that it represents the way a single participant behaves in that process. To borrow his example, when John moves his pen, he participates in the process of John moving his pen — which has as participant not only John but also his pen — and he also participates in the associated processual role that has as participant John but not his pen.[9]

Let us now go back to the example of John's history. Recall that John's disposition $d_{John}$ to move something is realized in the process $p_{move}$ of John moving his pen. From the perspective of processual roles, we can think of the process $p'_{move}$ of John moving *simpliciter* which is part of $p_{move}$ and which has as participant John but not his pen. Assume the PARTHOOD model of dispositions (introduced in Section 2.2). Since $d_{John}$ is realized in $p_{move}$, $d_{John}$ is also realized in $p'_{move}$ because, given the PARTHOOD model, a part of a realization of a disposition is also a realization of this disposition. Then, $p'_{move}$ is part of John's history, as it occupies the spatiotemporal region occupied by John. Hence, $d_{John}$ is realized in a process (namely $p'_{move}$) that is part of John's history.

In summary, it is *not* the case that the history of an independent continuant is the sum of *all* realizations of dispositions that inhere in any (proper or improper) part of the independent continuant during its whole life, as is shown by $d_{John}$ and $p_{move}$ in our example of John's moving his pen. We may hypothesize however that there is a subset of realizations (e.g. $p'_{move}$) of dispositions that inhere in parts of the independent continuant during its existence whose sum is the history of the independent continuant.

### 5.4. Evaluation

The causal approach can specify the identity of processes more directly than the compositional approach. However, it is committed to the potentially controversial these that any process is a realization of some disposition. To see the difficulty of this thesis, consider temporally discontinuous processes such as "my today eating process" in which I had breakfast in the morning, lunch in the afternoon, and dinner in the evening. We might hypothesize that such a discontinuous process is a realization of a single disposition to eat. Similarly, the mereological sum of the parts of a concert before and after the intermission might be analyzed as a realization of the disposition of the orchestra to play. Or consider a conference running over several days (namely, what happens during the conference itself, excluding the breaks to eat, sleep, etc.): this might be seen as a realization of the disposition of the agents participating in the conference to give talks, raise questions, provide responses, etc. A mereological theory of dispositions [13] would be useful to characterize such complex dispositions.

## 6. Discussion

We will discuss the alleged problem of too many processes (Section 6.1), a possible reduction of spatial changes to specifically dependent continuant changes (Section 6.2), and a possible reduction of indirect qualitative changes to direct qualitative changes (Section 6.3).

### 6.1. Too many processes?

One might worry that this view would lead to a too large number of processes. Indeed, for every process that extends over a temporally interval, there exists a different sub-process on each sub-interval of time. In case every time interval has an infinity of sub-intervals, this implies that an infinity of such sub-processes exists. Consider for example a process of John walking during time interval $i_1$. Then there exists a process of John walking during the first half of $i_1$, of him walking during the second fifth of $i_1$, of him walking during the 12$^{th}$ sixteenth of $i_1$, etc. However, this does not lead for us to any problematic form of multiplicativism, as those sub-processes are parts of the larger process — in the same way that a material entity may be composed of many (maybe an infinity of) material entities.

---

[9] Loebe [30] explains: "When John moves his pen, he and the pen form participants of that process, and the processual role which John plays captures what John does in that participation. Thinking of a mime who moves an imaginary pen should be a good illustration of the notion of a processual role" (ibid., p. 135).

## 6.2. A possible reduction of spatial changes to specifically dependent continuant changes

We distinguished two kinds of simple processes: specifically dependent continuant changes and spatial changes. We could think however that *Spatial change* is a subtype of *Specifically dependent continuant change* on an auxiliary assumption. According to Barton et Ethier's [31] ontological analysis of the term "velocity", an object-velocity is a disposition of the moving object to move. The ontology of the object-velocity could enable a spatial change of an independent continuant to be interpreted as a process that is a change of its object-velocity, on the condition that we would add (as GBG do) the notion of "stative change" when the specifically dependent continuant of an independent continuant does not change (to account for the case of a uniform motion process, where the object-velocity of the moving entity does not change).

## 6.3. A possible reduction of indirect qualitative changes to direct qualitative changes

Let us now explain how we can deal with indirect change in the sense defined by GBG as merely SDC change. Consider **apple$_0$**, which is green at t$_1$. That is, the skin of **apple$_0$** (which we will call **skin$_0$**) is green. This means that there is a quality **color_s$_0$** that inheres in **skin$_0$** and that instantiates the universal *Green* at t$_1$. However, in such situations, we often speak more simply of "the color of **apple$_0$**". This could be understood as implying the existence of a quality **color_a$_0$** that would inhere in **apple$_0$**. Here too, it instantiates the universal *Green* at t$_1$. Then, **color_a$_0$** and **color_s$_0$** are strongly related: in a sense, they reflect the same portion of reality (assuming for simplicity that the skin of an apple cannot be removed from the apple), and they always instantiate the same determinate universal of the determinable *Color*. This means that when the apple becomes red at t$_2$, both **color_a$_0$** and **color_s$_0$** instantiate the universal *Red*. However, as the former inheres in **apple$_0$** and the latter in **skin$_0$**, they cannot be identical. Therefore, if we accept that both the apple and its skin have a color, and that a quality inheres in only one bearer, we seem to be committed to the following informal "Principle of Quality Expansion" (on the model of Lombard's [32] Principle of Event Expansion, analyzed by GBG) or "PQE":

> **Principle of Quality Expansion** (PQE)
> If an independent continuant *x* has as part *y*, then: for any quality *q* of *y*, there is a quality *q'* of *x* such that *q* and *q'* correspond to the same portion of reality. (In particular, *q* changes whenever *q'* changes.)

We could elucidate the term "correspond to the same portion of reality" by means of truthmakers [33]: something in virtue of which a proposition is true (where the term "proposition" can be intuitively understood, its ontological nature being left aside).

If we accept the PQE, we can make sense of both direct and indirect qualitative change (in the sense of GBG) as simple SDC changes in BFO: what they would analyze as the direct qualitative change <**skin$_0$**, **color_s$_0$**, **t$_1$**> would correspond to our SDC change of **color_s$_0$**, whereas what they would analyze as the indirect qualitative change <**apple$_0$**, **color_s$_0$**, **t$_1$**> would correspond to our SDC change of **color_a$_0$**.

Note that this way to represent indirect changes is optional to our proposal: one might refuse to duplicate **color_s$_0$** into **color_a$_0$** and only accepts that the apple's skin — not the apple — has a color. In that case, one might speak of direct qualitative change and indirect qualitative change as GBG do, and refrain from accepting the entity **color_a$_0$**. However, by duplicating the quality of the color of the apple, we manage to reduce all qualitative changes to a same kind of SDC change. There is thus a trade-off between the number of introduced entities (e.g. in the apple scenario, two color qualities corresponding to the same reality vs. one) and the number of endorsed kinds of changes (only one kind of SDC change vs. both direct and indirect SDC changes). This view has another advantage insofar as it arguably accounts better for existing practices, as ontologies often consider qualities such as the color of an apple — even if it is more fundamentally a part of the apple (its skin) that is responsible for its color.

## 7. Related work

There is a huge body of philosophical literature on processes (or events, which may be a term more frequently used in philosophy). Although its comprehensive survey (e.g.[34]) is outside the purview of this article, there are two prominent views of them often called the "coarse-grained view" and the "fine-grained view". One typical version of the coarse-grained view says that processes are identical iff they occupy the same spatiotemporal region [28][29] — which we formalized as C1 and critically examined in Section 5.2. The fine-grained view, by contrast, characterizes the identity of processes in terms of properties in their broad sense (whether universals or particulars), as is illustrated by Kim's [35] view of processes as property exemplifications. By centering around an ontology of specifically dependent continuants such as dispositions, both compositional and causal approaches to the identity of processes we proposed can naturally belong to the group of the fine-grained view. It is worth remarking that the early Davidson [18] proposes a causal criterion for the identity of processes ("Events are identical iff they have the same causes and effects") and we may have proposed a dispositional version of such causal criterion.

In formal ontology, different upper ontologies develop different ontologies of processes and events (see e.g. Rodrigues & Abel's [36] general review). Guarino et al. ("GBG") [8] provide arguably one of the most systematic and general ontological analyses of events; indeed, we leveraged key elements of their work in developing a compositional approach to the identity of processes in Section 4. An alternative, considerably different view of processes and events from BFO's is that processes are mutable temporally extended entities and thus do change themselves while events are immutable temporally extended entities and thus do not change [37] (cf. [38] from a philosophical perspective). Events in this twofold ontology of occurrents would correspond to processes in BFO, while processes therein have no current equivalent in BFO. There are also many other views of the distinction between processes and events. To take just a few examples: processes are continuants rather than occurrents such as events [39][40]; processes are patterns of occurrence, whose concrete realizations may be viewed as events or states [41]; and processes are physical entities, whereas events are mental and social entities [42][43][44].

## 8. Conclusion and future work

We investigated the identity of processes with a focus on the BFO category of process. The resulting two approaches are the compositional approach that is based on two simple kinds of processes (specifically dependent continuant changes and spatial changes) and the causal approach that is based on a dispositional view of processes. In the future we will further each of these two approaches. As for the compositional approach, we will scrutinize the PDH based on the conclusion for it that is given in Section 4.5.4. As for the causal approach, it is worth investigating the relationship between the identity of processes and the identity of dispositions [14]. An important question will be whether the compositional and causal criteria lead to the same results concerning the identity of processes or not. Our long-term goal is to integrate both approaches so as to develop a systematic theory of the identity of processes, in the hope that the resulting theory will help to clarify various process-related entities such as process profiles [5] — whose introduction is motivated to explain the same aspect of different processes — and states (for initial thoughts, see Galton's [41] discussion that the term "state" may refer to two different entities: a continuant entity and an occurrent entity).

## 9. Acknowledgements

We benefited from interesting discussions on related topics: with Nicola Guarino and Riccardo Baratella on their theory of events and qualities, and with Alan Ruttenberg on substance creation and destruction in BFO. Fumiaki Toyoshima is financially supported by the Japan Society for the Promotion of Science (JSPS).

## 10. References

[1] Ashburner M, Ball CA, Blake JA, Botstein D, Butler H, Cherry JM, Davis AP, Dolinski K, Dwight SS, Eppig JT, Harris MA, Hill DP, Issel-Tarver L, Kasarskis A, Lewis S, Matese JC, Richardson JE, Ringwald M, Rubin GM, Sherlock G. Gene ontology: tool for the unification of biology. The Gene Ontology Consortium. Nat Genet. 2000 May;25(1):25-9. doi: 10.1038/75556.

[2] Arp R, Smith B, Spear AD. Building ontologies with Basic Formal Ontology. MIT Press; 2015. 248 p.

[3] ISO/IEC PRF 21838-2.2 [internet]. Information technology - Top-level ontologies (TLO) - Part 2: Basic Formal Ontology (BFO); 2020 Mar. URL: https://www.iso.org/standard/74572.html

[4] Otte JN, Beverley J, and Ruttenberg, A. BFO: Basic Formal Ontology. Appl Ontol. 2022 Jan;15(1):17-43. doi: 10.3233/AO-220262

[5] Smith B. Classifying processes: an essay in applied ontology. Ratio. 2012;25:463-88.

[6] Jarrar M, Ceusters W. Classifying processes and Basic Formal Ontology. In: Proceedings of ICBO2017. CEUR Workshop Proceedings, vol. 2137; 2017. p. 1-6.

[7] Garbacz P. A new perspective on criteria of identity. Synthese 200, 476 (2022). https://doi.org/10.1007/s11229-022-03954-x

[8] Guarino N, Baratella R, Guizzardi G. Events, their names, and their synchronic structure. Appl Ontol. 2022 May;17(2):249-83. doi: 10.3233/AO-220261

[9] Smith B, Ceusters W. Ontological realism: a methodology for coordinated evolution of scientific ontologies. Appl Ontol. 2010 Nov;5(3-4):139-88. doi: 10.3233/AO-2010-0079

[10] Toyoshima F, Barton A, Jansen L, Ethier, JF. Towards a unified dispositional framework for realizable entities. In: Proceedings of FOIS2021. Amsterdam: IOS Press; 2018. p. 13-7. doi: 10.3233/FAIA210371

[11] Galton A. The treatment of time in upper ontologies. In: Proceedings of FOIS2018. Amsterdam: IOS Press; 2018. p. 33-46. doi: 10.3233/978-1-61499-910-2-33

[12] Röhl J, Jansen L. Representing dispositions. J Biomed Semant. 2011 Aug;2(Suppl 4):S4.

[13] Barton A, Jansen L, Ethier JF. A taxonomy of disposition-parthood. In: Proceedings of JOWO2017. CEUR Workshop Proceedings, vol. 2050; 2017. p. 1-10.

[14] Barton A, Grenier O, Jansen L, Ethier JF. The identity of dispositions. In: Proceedings of FOIS2018. Amsterdam: IOS Press; 2018. p. 113-25. doi: 10.3233/978-1-61499-910-2-113

[15] Varzi AC, Cotnoir AJ. Mereology. Oxford: Oxford University Press; 2021. 424 p.

[16] Masolo C, Borgo S, Gangemi A, Guarino N, Oltramari A. Wonderweb deliverable D18 - ontology library (final). LOA-NCR-ISTC; 2003.
URL: http://wonderweb.man.ac.uk/deliverables/D18.shtml

[17] Borgo S, Ferrario R, Gangemi A, Guarino N, Masolo C, Porello D, Sanfilippo EM, Vieu, L. DOLCE: A Descriptive Ontology for Linguistic and Cognitive Engineering. Appl Ontol. 2022 Jan;15(1):45-69. doi: 10.3233/AO-210259

[18] Davidson D. The individuation of events. In: N Rescher, editor. Essays in honor of Carl G. Hempel. Dordrecht: Reidelpp; 1969:216-34.

[19] Wilson J. Determinables and determinates. In: Zalta EN, Nodelman U, editors. The Stanford Encyclopedia of Philosophy. Spring 2023 ed.
URL: https://plato.stanford.edu/archives/spr2023/entries/determinate-determinables/

[20] van Riel R, Van Gulick R. Scientific reduction. In: Zalta EN, editor. The Stanford Encyclopedia of Philosophy. Spring 2019 ed.
URL: https://plato.stanford.edu/archives/spr2019/entries/scientific-reduction/

[21] Tutorial on Basic Formal Ontology. Last edited on February 9, 2019.

URL: https://ncorwiki.buffalo.edu/index.php/Tutorial_on_Basic_Formal_Ontology
[22] Ceusters W, Smith B. Aboutness: towards foundations for the Information Artifact Ontology. In: Proceedings of ICBO2015. CEUR Workshop Proceedings, vol. 1515; 2015. p. 1-5.
[23] Barton A, Toyoshima F, Vieu L, Fabry P, Ethier JF. The mereological structure of informational entities. In: Proceedings of FOIS2020. Amsterdam: IOS Press; 2020. p. 201-15. doi: 10.3233/FAIA200672
[24] Brochhausen M, Almeida MB, Slaughter L. Towards a formal representation of document acts and the resulting legal entities. In: Svennerlind S, Almäng J, Ingthorsson R, editors. Johanssonian investigations: essays in honour of Ingvar Johansson on his seventieth birthday. Walter de Gruyter; 2013. p. 120-139.
[25] Mumford S Anjum, RL. Getting causes from powers. Oxford University Press; 2011. 272 p.
[26] Williams N. The powers metaphysic. Oxford University Press; 2019. 268 p.
[27] Barton A, Rovetto R, Mizoguchi R. Newtonian forces and causation: a dispositional account. In: Proceedings of FOIS2014. Amsterdam: IOS Press; 2014. p. 157-70. doi: 10.3233/978-1-61499-438-1-157
[28] Quine WVO. Events and reification. In: Lepore E, Mc Laughlin B, editors. Actions and events: perspectives on the philosophy of Donald Davidson. Oxford: Basil Blackwell; 1985:162-71.
[29] Davidson D. Reply to Quine on events. In: Lepore E, Mc Laughlin B, editors. Actions and events: perspectives on the philosophy of Donald Davidson. Oxford: Basil Blackwell; 1985:172-76.
[30] F. Loebe. Abstract vs. social roles - Towards a general theoretical account of roles. Appl Ontol. 2007;2(2):127-58.
[31] Barton A, Ethier JF. The two ontological faces of velocity. In: Proceedings of FOIS2016. Amsterdam: IOS Press; 2016. p. 123-36. doi: 10.3233/978-1-61499-660-6-123
[32] Lombard LB. Events: a metaphysical study. Routledge; 1986. 272 p.
[33] MacBride F. Truthmakers. In: Zalta EN, Nodelman U, editors. The Stanford Encyclopedia of Philosophy. Fall 2022 ed.
URL: https://plato.stanford.edu/archives/fall2022/entries/truthmakers/
[34] Casati R, Varzi A. Events. In: Zalta EN, Nodelman U, editors. The Stanford Encyclopedia of Philosophy. Summer 2020 ed.
URL: https://plato.stanford.edu/archives/sum2020/entries/events/
[35] Kim J. Events as property exemplifications. In: Brand M, Walton D, editors. Action theory. Dordrecht: Reidel; 1976. p. 159-77.
[36] Rodrigues FH, Abel M. What to consider about events: A survey on the ontology of occurrents. Appl Ontol. 2019;14(4):343-78. doi: 10.3233/AO-190217
[37] Galton A, Mizoguchi R. The water falls but the waterfall does not fall: new perspectives on objects, processes and events. Appl Ontol. 2009;4(2):71-107. doi:10.3233/AO-2009-0067
[38] Stout R, editor. Process, action, and experience. Oxford University Press. 2018. 240 p.
[39] Stout R. Processes. Philosophy. 1997;72(279):19-27.
[40] Galton A. On what goes on: the ontology of processes and events. In: Proceedings of FOIS2006. Amsterdam: IOS Press; 2006. p. 4-11.
[41] Galton A. Processes as patterns of occurrence. In [38]. p. 41-57.
[42] Gill K. On the metaphysical distinction between processes and events. Can. J. Philos. 1993 Sep;23(3):365-84.
[43] Kassel, G. Processes endure, whereas events occur. In: Borgo S, Ferrario R, Masolo C, Vieu L, editors. Ontology makes sense: essays in honor of Nicola Guarino. Amsterdam: IOS Press; 2019. p. 177-93.
[44] Kassel, G. Physical processes, their life and their history. Appl Ontol. 2020 May;15(2):109-33. doi: 10.3233/AO-200222.